\documentclass[preprint,showpacs,preprintnumbers,amsmath,amssymb]{revtex4}

\usepackage{graphicx}
\usepackage{dcolumn}
\usepackage{bm}

\begin{document}

\preprint{}

\title{Third-order integrable difference equations generated by a pair of second-order equations}

\author{Junta Matsukidaira}
\email{junta@math.ryukoku.ac.jp}
\affiliation{%
Department of Applied Mathematics and Informatics, 
Ryukoku University, Seta, Otsu, Shiga 520-2194, Japan
}%

\author{Daisuke Takahashi}
\email{daisuket@waseda.jp}
\affiliation{%
Department of Mathematical Sciences,\\
Waseda University, 3-4-1, Ohkubo, Shinjuku-ku, Tokyo 169-8555, Japan
}%

\date{\today}

\begin{abstract}
We show that the third-order difference equations proposed by Hirota,
 Kimura and Yahagi are generated by a pair of second-order difference
 equations. In some cases, the pair of the second-order equations are
 equivalent to the Quispel-Robert-Thomson(QRT) system, but in the other
 cases, they are irrelevant to the QRT system. We also discuss
 an ultradiscretization of the equations.

\end{abstract}

\pacs{02.30Ik, 05.45.-a}
\maketitle
\section{Introduction}

Discrete integrable systems have attracted much attention and a lot of
studies have been done by various points of view, such as integrability
criteria (singularity confinement property\cite{GRP}, algebraic
entropy\cite{HV}), geometric or algebraic description of the
equations\cite{Sakai, Take, RGO1, RGO2, Tsuda, KMNOY} and so on.

Especially, second-order integrable difference equations including QRT
system\cite{QRT1,QRT2} and discrete Painlev\'e equations\cite{RGH} which
are regarded as non-autonomous variations of QRT system, have been
extensively studied, and a number of significant properties have been
obtained.

For example, a symmetric version of QRT system is defined by the
following form
\begin{equation}
 x_{n+1} = \frac{f_1(x_n) - x_{n-1}f_2(x_n)}{f_2(x_n) - x_{n-1}f_3(x_n)},
\end{equation}
where $f_j(x)$ is defined by
\begin{equation}
 \begin{pmatrix}
  f_1(x)\\
  f_2(x)\\
  f_3(x)\\
 \end{pmatrix}
 = A
 \begin{pmatrix}
  x^2 \\
  x   \\
  1   \\
 \end{pmatrix}
 \times
 B
 \begin{pmatrix}
  x^2 \\
  x   \\
  1   \\
 \end{pmatrix},
\end{equation}
with arbitrary symmetric 3$\times$3 matrices $A$ and $B$. This equation
has a conserved quantity $h(x_{n-1}, x_n)$ defined by $A$ and
$B$. Moreover, a general solution is described by an elliptic function.

However very few results have been obtained for third-order integrable
difference equations. Research on such equations are important in order
to reveal integrable structures of general discrete integrable systems.

In this paper, we investigate third-order integrable difference
equations proposed by Hirota, Kimura and Yahagi\cite{HKY} and show that
they are generated by a pair of second-order integrable difference
equations. Moreover we also discuss their ultradiscretization.

\begin{figure}[htbp]
 \includegraphics{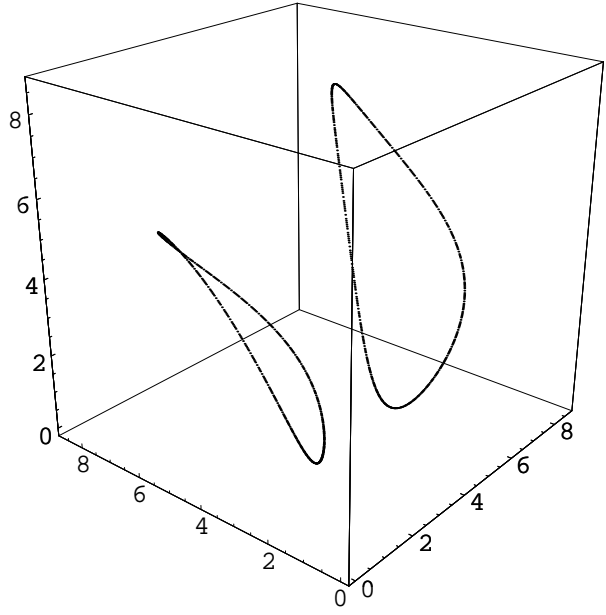} \caption{ \label{fig:fig1}A trajectory of a
 solution to Eq.~\eqref{mY6} for $a=2.0,\, y_0=1.0,\, y_1=2.0,\,
 y_2=1.5$}
\end{figure}

Hirota, Kimura and Yahagi have investigated third order difference
equations of the form
\begin{equation}
 x_{n+2}x_{n-1} = \frac{a_0+a_1x_n+a_2x_{n+1}+a_3x_nx_{n+1}}{b_0+b_1x_n+b_2x_{n+1}+b_3x_nx_{n+1}},
\end{equation}
and have found that nine equations, 
\begin{align}
 x_{n+2}x_{n-1} &=
  \frac{a_0+a_1(x_n+x_{n+1})+a_3x_nx_{n+1}}{a_3+b_1(x_n+x_{n+1})+b_3x_nx_{n+1}},
 \tag{Y1} \label{Y1}\\
 x_{n+2}x_{n-1} &=
 \frac{a_0(1+x_n+x_{n+1})+a_3x_nx_{n+1}}{a_0+a_3(x_n+x_{n+1}+x_nx_{n+1})}, \tag{Y2} \label{Y2}\\
 x_{n+2}x_{n-1} &=
 \frac{a_0(-1+x_n-x_{n+1})+a_3x_nx_{n+1}}{a_0+a_3(x_n-x_{n+1}-x_nx_{n+1})}, \tag{Y3} \label{Y3}\\
 x_{n+2}x_{n-1} &=
 \frac{a_0+a_1(x_n+x_{n+1}+x_nx_{n+1})}{1+x_n+x_{n+1}+x_nx_{n+1}},
 \tag{Y4} \label{Y4}\\ x_{n+2}x_{n-1} &=
 \frac{a_1(x_n-x_{n+1})+a_3x_nx_{n+1}}{a_3+b_1(-x_n+x_{n+1})}, \tag{Y5} \label{Y5}\\
 x_{n+2}x_{n-1} &= \frac{a_3x_nx_{n+1}}{b_1(x_n+x_{n+1})+b_3x_nx_{n+1}},
 \tag{Y6} \label{Y6}\\
 x_{n+2}x_{n-1} &= \frac{a_0+a_1x_n}{a_1x_n+a_0x_nx_{n+1}}, \tag{Y7} \label{Y7}\\ 
 x_{n+2}x_{n-1} &= \frac{a_0+a_1x_n}{-a_1x_n+a_0x_nx_{n+1}}, \tag{Y8} \label{Y8}\\ 
 x_{n+2}x_{n-1} &= \frac{x_n+x_nx_{n+1}}{1+x_n}, \tag{Y9} \label{Y9}
\end{align}
are integrable in the sense that they have two independent conserved
quantities. A remarkable property of these equations is that their
trajectory of a solution in 3-dimensional phase space looks like a
composition of two separate curves. Fig.~\ref{fig:fig1} is an example of
such trajectories in 3D phase space which is generated by
\begin{equation}\label{mY6}
 y_{n+2}y_{n-1} = a + y_n + y_{n+1},
\end{equation}
where the equation is obtained through a variable transformation $y_n =
\dfrac{a_3}{b_1x_n},\ a = \dfrac{a_3b_3}{b_1^2}$ from Eq.~\eqref{Y6}.
Moreover, it is an important fact that odd step points belong to the one
curve and even step points belong to the other.

This fact strongly suggests that a combination of lower dimensional
integrable equations determines the integrability of the third-order
difference equation. We show this is true for all nine equations in the
following section.

\section{pair of second-order integrable equations generating a third-order equation}
\subsection{Y6}
If we take a backward difference of Eq.~\eqref{mY6}
\begin{equation*}
 y_{n+2}y_{n-1} = a + y_n + y_{n+1}, 
\end{equation*}
we obtain
\begin{equation}\label{dmY6}
 \Delta_n(y_{n+2}y_{n-1} - a - y_n - y_{n+1}) = 0,
\end{equation}
where $\Delta_n$ is a difference operator defined by $\Delta_n
f_n=f_n-f_{n-1}$. Eq.~\eqref{dmY6} can be written as
\begin{equation}
 \dfrac{(1+y_{n+2})(1+y_n)}{y_{n+1}}
 = \dfrac{(1+y_n)(1+y_{n-2})}{y_{n-1}}.
\end{equation}
This formula means that there are constants which depend on the
initial values and on a parity of $n$. Hence, we obtain
\begin{equation}\label{Y6c}
\begin{cases}
  \dfrac{(1+g_{n+1})(1+g_n)}{h_n}
  &= c_0,\\
 \vbox to 8mm{}
  \dfrac{(1+h_n)(1+h_{n-1})}{g_n}
  &= c_1,
\end{cases}
\end{equation}
where $g_n = y_{2n},\ h_n = y_{2n+1},\ c_0 =
\dfrac{(1+y_0)(1+y_2)}{y_1},\ c_1=\dfrac{(1+y_1)(1+y_3)}{y_2}$. From
Eqs.~\eqref{mY6} and \eqref{Y6c}, we obtain a pair of second-order
difference equations
\begin{align}
  g_{n+1} &= \dfrac{(1+ac_0)+(1+c_0)g_n}{g_{n-1}(1+g_n)}, \label{Y6g}\\
  h_{n+1} &= \dfrac{(1+ac_1)+(1+c_1)h_n}{h_{n-1}(1+h_n)}, \label{Y6h}
\end{align}
where Eq.~\eqref{Y6g} is a equation for even steps and Eq.~\eqref{Y6h}
is a equation for odd steps respectively. Eq.~\eqref{Y6g} and
Eq.~\eqref{Y6h} can be written in the QRT form
\begin{equation}
\left\{
\begin{aligned}
  g_{n+1} &= \frac{G_1(g_n)-g_{n-1}G_2(g_n)}{G_2(g_n)-g_{n-1}G_3(g_n)}, \\
\vbox to 8mm{}
  h_{n+1} &= \frac{H_1(h_n)-h_{n-1}H_2(h_n)}{H_2(h_n)-h_{n-1}H_3(h_n)},
\end{aligned}
\right.
\end{equation}
where
\begin{align}
 A(c) &= 
   \begin{pmatrix}
    1     & 2+c & 1+c\\
    2+c & 0     & 2+2c+ac+c^2\\
    1+c & 2+2c+ac+c^2 & (1+c)(1+ac)
   \end{pmatrix},\qquad
 B = 
   \begin{pmatrix}
    0 & 0 & 0 \\
    0 & 1 & 0 \\
    0 & 0 & 0
   \end{pmatrix},
\end{align}
\begin{align}
 \begin{pmatrix}
  G_1(x) \\
  G_2(x) \\
  G_3(x) \\
 \end{pmatrix}
 &= A(c_0)
 \begin{pmatrix}
  x^2\\
  x\\
  1
 \end{pmatrix} \times
 B
 \begin{pmatrix}
  x^2\\
  x\\
  1
 \end{pmatrix},\qquad
 \begin{pmatrix}
  H_1(x) \\
  H_2(x) \\
  H_3(x) \\
 \end{pmatrix}
 = A(c_1)
 \begin{pmatrix}
  x^2\\
  x\\
  1
 \end{pmatrix}\times
 B
 \begin{pmatrix}
  x^2\\
  x\\
  1
 \end{pmatrix}.
\end{align}
Consequently, conserved quantities of Eqs.~\eqref{Y6g} and \eqref{Y6h}
are given as
\begin{align}
 k_0 = & \Bigl(g_n^2g_{n+1}^2 + (2+c_0)(g_n+g_{n+1})g_ng_{n+1} +
 (1+c_0)(g_n^2 + g_{n+1}^2)  \nonumber\\
 & + (2+2c_0+ac_0+c_0^2)(g_n + g_{n+1}) + (1+c_0)(1+ac_0)\Bigr)
 /(g_ng_{n+1}),
\end{align}
and
\begin{align}
 k_1 = & \Bigl(h_n^2h_{n+1}^2 + (2+c_1)(h_n+h_{n+1})h_nh_{n+1} +
 (1+c_1)(h_n^2 + h_{n+1}^2)  \nonumber\\
 & + (2+2c_1+ac_1+c_1^2)(h_n + h_{n+1}) + (1+c_1)(1+ac_1)\Bigr)
 /(h_nh_{n+1}).
\end{align}
Hence invariant curves of Eqs.~\eqref{Y6g} and \eqref{Y6h} are given by
the above equations. These curves determine the structure of the
trajectory of a solution to Eq.~\eqref{Y6} in 3D phase space shown in
Fig.~\ref{fig:fig1}. This is the simplified integrability structure of
Eq.~\eqref{Y6} and we show below a similar structure exists in the other
eight equations.

\subsection{Y1}
From Eq.~\eqref{Y1}
\begin{equation}
 x_{n+2}x_{n-1} =
  \frac{a_0+a_1(x_n+x_{n+1})+a_3x_nx_{n+1}}{a_3+b_1(x_n+x_{n+1})+b_3x_nx_{n+1}},\tag{Y1}
\end{equation}
we obtain
\begin{equation}\label{dY1}
 \Delta_n( (a_3+b_1(x_n+x_{n+1})+b_3x_nx_{n+1})x_{n+2}x_{n-1}-
  (a_0+a_1(x_n+x_{n+1})+a_3x_nx_{n+1}) )=0.
\end{equation}
Eq.~\eqref{dY1} can be written as
\begin{align}
& b_3x_{n+2}x_n+b_1\left(x_{n+2}+x_n+\frac{x_{n+2}x_n}{x_{n+1}}\right)+a_3\left(\frac{x_{n+2}}{x_{n+1}}+\frac{x_{n}}{x_{n+1}}\right)+\frac{a_1}{x_{n+1}}
\nonumber\\
&= b_3x_nx_{n-2}+b_1\left(x_n+x_{n-2}+\frac{x_nx_{n-2}}{x_{n-1}}\right)+a_3\left(\frac{x_n}{x_{n-1}}+\frac{x_{n-2}}{x_{n-1}}\right)+\frac{a_1}{x_{n-1}}.
\end{align}
Hence we obtain
\begin{equation}
 \begin{cases}\label{Y1c}
  g_n = \dfrac{a_1 + a_3(h_{n-1} + h_n) +
  b_1h_{n-1}h_n}{c_1-b_1(h_{n-1}+h_n)-b_3h_{n-1}h_n},\\
  \vbox to 8mm{}
  h_n = \dfrac{a_1 + a_3(g_n + g_{n+1}) +
  b_1g_ng_{n+1}}{c_0-b_1(g_n+g_{n+1})-b_3g_ng_{n+1}},
 \end{cases}
\end{equation}
where $g_n = x_{2n},\ h_n = x_{2n+1}$ and
\begin{equation}
 \begin{split}
  c_0 &= \dfrac{1}{x_1}(b_1x_0x_2 + a_3x_2 + a_3x_0 + a_1) +
  b_3x_0x_2 + b_1(x_0 + x_2),\\
  c_1 &= \dfrac{1}{x_2}(b_1x_1x_3 + a_3x_3 + a_3x_1 + a_1) + b_3x_1x_3 + b_1(x_1 + x_3).
 \end{split}
\end{equation}
From Eqs.~\eqref{Y1} and \eqref{Y1c}, we obtain a pair of
QRT systems
\begin{equation}
\left\{
\begin{aligned}
  g_{n+1} &= \frac{G_1(g_n)-g_{n-1}G_2(g_n)}{G_2(g_n)-g_{n-1}G_3(g_n)}, \\
\vbox to 8mm{}
  h_{n+1} &= \frac{H_1(h_n)-h_{n-1}H_2(h_n)}{H_2(h_n)-h_{n-1}H_3(h_n)},
\end{aligned}
\right.
\end{equation}
where
\begin{align}
 A(c) &= 
  \begin{pmatrix}
   b_1^2 - 2a_3b_3 + \dfrac{a_1b_3^2}{b_1}  & 0 & a_3^2 - a_1b_1 \\
   0 & 2a_3^2-a_0b_3+2a_3c-\dfrac{a_1b_3c}{b_1} &
   2a_1a_3-a_0b_1+a_1c \\
   \\
   a_3^2-a_1b_1 & 2a_1a_3-a_0b_1+a_1c & a_1^2+a_0c
  \end{pmatrix},\\
 B(c) &= 
 \begin{pmatrix}
  b_3 & b_1 &  0 \\
  b_1 & -c  & 0 \\
  0 & 0 & 0 \\
 \end{pmatrix},
\end{align}
\begin{equation}
 \begin{pmatrix}
  G_1(x) \\
  G_2(x) \\
  G_3(x) \\
 \end{pmatrix}
 = A(c_0)
 \begin{pmatrix}
  x^2\\
  x\\
  1
 \end{pmatrix}\times
 B(c_0)
 \begin{pmatrix}
  x^2\\
  x\\
  1
 \end{pmatrix},\qquad
 \begin{pmatrix}
  H_1(x) \\
  H_2(x) \\
  H_3(x) \\
 \end{pmatrix}
 = A(c_1)
 \begin{pmatrix}
  x^2\\
  x\\
  1
 \end{pmatrix}\times
 B(c_1)
 \begin{pmatrix}
  x^2\\
  x\\
  1
 \end{pmatrix}
\end{equation}

\subsection{Y4}
From Eq.~\eqref{Y4}
\begin{equation}
  x_{n+2}x_{n-1}=\frac{a_0+a_1(x_{n+1}+x_n+x_{n+1}x_n)}{1+x_n+x_{n+1}+x_{n+1}x_n}, \tag{Y4}
\end{equation}
we obtain
\begin{equation}\label{dY4}
 \Delta_n((1+x_n+x_{n+1}+x_{n+1}x_n)x_{n+2}x_{n-1}-(a_0+a_1(x_{n+1}+x_n+x_{n+1}x_n)))
  = 0.
\end{equation}
Eq.~\eqref{dY4} can be written as
\begin{equation}
 (x_{n+2}+a_1)(x_n+a_1)\frac{x_{n+1}+1}{x_{n+1}}=(x_n+a_1)(x_{n-2}+a_1)\frac{x_{n-1}+1}{x_{n-1}}.
\end{equation}
From this equation, we obtain
\begin{equation}\label{Y4c}
 \begin{cases}
  g_n=\dfrac{(h_n+a_1)(h_{n-1}+a_1)}{c_1-(h_n+a_1)(h_{n-1}+a_1)},\\
  \vbox to 8mm{}
  h_n=\dfrac{(g_{n+1}+a_1)(g_n+a_1)}{c_0-(g_{n+1}+a_1)(g_n+a_1)},
 \end{cases}
\end{equation}
where $g_n = x_{2n},\ h_n = x_{2n+1}$ and
\begin{align}
 c_0 &= (x_2+a_1)(x_0+a_1)\dfrac{x_1+1}{x_1},\\
 c_1 &= (x_3+a_1)(x_1+a_1)\dfrac{x_2+1}{x_2}.
\end{align}
From Eqs.~\eqref{Y4} and \eqref{Y4c}, we obtain a pair of
QRT systems
\begin{equation}
\left\{
\begin{aligned}
  g_{n+1} &= \frac{G_1(g_n)-g_{n-1}G_2(g_n)}{G_2(g_n)-g_{n-1}G_3(g_n)}, \\
\vbox to 8mm{}
  h_{n+1} &= \frac{H_1(h_n)-h_{n-1}H_2(h_n)}{H_2(h_n)-h_{n-1}H_3(h_n)},
\end{aligned}
\right.
\end{equation}
where
\begin{equation}
 \begin{split}
  A(c) &= 
  \begin{pmatrix}
   (a_0-a_1-c)(a_1-1) & 0 & -(a_0-a_1-c)(a_1-1)a_1^2 \\
   0      & a_{22} & a_{23} \\
   -(a_0-a_1-c)(a_1-1)a_1^2 & a_{32} & a_{33} \\
  \end{pmatrix},\\
  a_{22} &= -2(a_0-a_1)(a_1-1)a_1^2 + c(a_0-2a_1+a_1^3-c), \\
  a_{23} &= -2(a_0-a_1)(a_1-1)a_1^3 + ca_1(-a_0+2a_0a_1-2a_1^2+a_1^3-a_1c), \\
  a_{32} &= a_{23},\\
  a_{33} &= -(a_0-a_1)(a_1-1)a_1^4 + ca_1(-a_0a_1+2a_0a_1^2-a_1^3-a_0c)
 \end{split}
\end{equation}
\begin{align}
 B(c) &= 
 \begin{pmatrix}
  1 & a_1 &  0 \\
  a_1 & a_1^2-c  & 0 \\
  0 & 0 & 0 \\
 \end{pmatrix},
\end{align}
\begin{equation}
 \begin{pmatrix}
  G_1(x) \\
  G_2(x) \\
  G_3(x) \\
 \end{pmatrix}
 = A(c_0)
 \begin{pmatrix}
  x^2\\
  x\\
  1
 \end{pmatrix}\times
 B(c_0)
 \begin{pmatrix}
  x^2\\
  x\\
  1
 \end{pmatrix},\qquad
 \begin{pmatrix}
  H_1(x) \\
  H_2(x) \\
  H_3(x) \\
 \end{pmatrix}
 = A(c_1)
 \begin{pmatrix}
  x^2\\
  x\\
  1
 \end{pmatrix}\times
 B(c_1)
 \begin{pmatrix}
  x^2\\
  x\\
  1
 \end{pmatrix}
\end{equation}

\subsection{Y5}
From Eq.~\eqref{Y5}
\begin{equation}
  x_{n+2}x_{n-1}=\frac{a_1x_n-a_1x_{n+1}+a_3x_nx_{n+1}}{a_3-b_1x_n+b_1x_{n+1}}. \tag{Y5}
\end{equation}
we obtain
\begin{equation}\label{dY5}
 \begin{split}
  & x_{n+2}x_{n-1}(a_3-b_1x_n+b_1x_{n+1})-(a_1x_n-a_1x_{n+1}+a_3x_nx_{n+1})\\  
  & +x_{n+1}x_{n-2}(a_3-b_1x_{n-1}+b_1x_n)-(a_1x_{n-1}-a_1x_n+a_3x_{n-1}x_n)
  = 0.
 \end{split}
\end{equation}
Note that we don't take a backward difference but a sum of Eq.~\eqref{Y5}
here. Eq.~\eqref{dY5} can be written as
\begin{equation}
  a_3\frac{x_{n+2}-x_n}{x_{n+1}}-\frac{a_1}{x_{n+1}}+b_1(x_{n+2}+x_n)-b_1\frac{x_{n+2}x_n}{x_{n+1}}= a_3\frac{x_n-x_{n-2}}{x_{n-1}}-\frac{a_1}{x_{n-1}}+b_1(x_n+x_{n-2})-b_1\frac{x_nx_{n-2}}{x_{n-1}}.
\end{equation}
From this equation, we obtain
\begin{equation}\label{Y5c}
 \begin{cases}
  g_n =
  \dfrac{a_3(h_n-h_{n-1})-a_1-b_1h_nh_{n-1}}{c_1-b_1(h_n+h_{n-1})},\\
  \vbox to 8mm{}
  h_n = \dfrac{a_3(g_{n+1}-g_n)-a_1-b_1g_{n+1}g_n}{c_0-b_1(g_{n+1}+g_n)}, \end{cases}
\end{equation}
where $g_n = x_{2n},\ h_n = x_{2n+1}$ and
\begin{align}
 c_0 &=
 a_3\frac{x_2-x_0}{x_1}-\frac{a_1}{x_1}+b_1(x_2+x_0)-b_1\frac{x_2x_0}{x_1},\\
 c_1 &= a_3\frac{x_3-x_1}{x_2}-\frac{a_1}{x_2}+b_1(x_3+x_1)-b_1\frac{x_3x_1}{x_2}.
\end{align}
From Eqs.~\eqref{Y5} and \eqref{Y5c}, we obtain a pair of
QRT systems
\begin{equation}
\left\{
\begin{aligned}
  g_{n+1} &= \frac{G_1(g_n)-g_{n-1}G_2(g_n)}{G_2(g_n)-g_{n-1}G_3(g_n)}, \\
\vbox to 8mm{}
  h_{n+1} &= \frac{H_1(h_n)-h_{n-1}H_2(h_n)}{H_2(h_n)-h_{n-1}H_3(h_n)},
\end{aligned}
\right.
\end{equation}
where
\begin{align}
 A(c) &= 
  \begin{pmatrix}
   b_1^2 & 0 & -a_3^2-a_1b_1 \\
   0     & 2a_3^2 & a_1c \\
   -a_3^2-a_1b_1 & a_1c & a_1^2
  \end{pmatrix},\\
 B(c) &= 
 \begin{pmatrix}
  0   & b_1 & 0 \\
  b_1 &  -c & 0 \\
  0   &   0 & 0 \\
 \end{pmatrix},
\end{align}
\begin{equation}
 \begin{pmatrix}
  G_1(x) \\
  G_2(x) \\
  G_3(x) \\
 \end{pmatrix}
 = A(c_0)
 \begin{pmatrix}
  x^2\\
  x\\
  1
 \end{pmatrix}\times
 B(c_0)
 \begin{pmatrix}
  x^2\\
  x\\
  1
 \end{pmatrix},\qquad
 \begin{pmatrix}
  H_1(x) \\
  H_2(x) \\
  H_3(x) \\
 \end{pmatrix}
 = A(c_1)
 \begin{pmatrix}
  x^2\\
  x\\
  1
 \end{pmatrix}\times
 B(c_1)
 \begin{pmatrix}
  x^2\\
  x\\
  1
 \end{pmatrix}.
\end{equation}

\subsection{Y7}
Introducing a variable transformation $x_n=\dfrac{f_{n+1}}{f_n}$ to
Eq.~\eqref{Y7}
\begin{equation}
   x_{n+2}x_{n-1}=\frac{a_0+a_1x_n}{a_1x_n+a_0x_nx_{n+1}}, \tag{Y7}
\end{equation}
we obtain
\begin{equation}
 a_0(f_{n+1}+f_{n-1})+a_1\frac{f_{n+1}f_{n-1}}{f_n} = a_0(f_{n+3}+f_{n+1})+a_1\frac{f_{n+3}f_{n+1}}{f_{n+2}}.
\end{equation}
From this equation, we obtain
\begin{equation}\label{Y7c}
 \begin{cases}
  a_0(g_{n+1}+g_n)+a_1\dfrac{g_{n+1}g_n}{h_n} =c_0\\
  \vbox to 8mm{}
  a_0(h_n+h_{n-1})+a_1\dfrac{h_nh_{n-1}}{g_n} =c_1
 \end{cases}
\end{equation}
where $g_n = f_{2n},\ h_n = f_{2n+1}$ and
\begin{align}
  c_0 &= a_0(g_1+g_0)+a_1\frac{g_1g_0}{h_0}=f_0a_0(1+x_0x_1+\frac{a_1}{a_0}x_1), \\
  c_1 &=
 a_0(h_1+h_0)+a_1\frac{h_1h_0}{g_1}=f_0a_0x_0(1+x_1x_2+\frac{a_1}{a_0}x_2).
\end{align}
From Eq.~\eqref{Y7c}, we obtain a pair of QRT system
\begin{equation}
\left\{
\begin{aligned}
  g_{n+1} &= \frac{G_1(g_n)-g_{n-1}G_2(g_n)}{G_2(g_n)-g_{n-1}G_3(g_n)}, \\
\vbox to 8mm{}
  h_{n+1} &= \frac{H_1(h_n)-h_{n-1}H_2(h_n)}{H_2(h_n)-h_{n-1}H_3(h_n)},
\end{aligned}
\right.
\end{equation}
where 
\begin{align}
 \begin{pmatrix}
  G_1(x) \\
  G_2(x) \\
  G_3(x) \\
 \end{pmatrix}
 = &
 \begin{pmatrix}
      0 & 0 & a_0^2(a_0^2-a_1^2)\\
      0 & (a_0^2-a_1^2)(2a_0^2-a_1^2) & -a_0(2a_0^2c_0 - a_1^2c_0 + a_0a_1c_1)\\
  a_0^2(a_0^2-a_1^2) & -a_0(2a_0^2c_0 - a_1^2c_0 + a_0a_1c_1) & a_0c_0(a_0c_0+a_1c_1)\\
 \end{pmatrix}
 \begin{pmatrix}
  x^2\\
  x\\
  1
 \end{pmatrix}\nonumber\\
 & \times
 \begin{pmatrix}
  a_0a_1 &  a_0c_1 & 0 \\
  a_0c_1 & -c_0c_1 & 0 \\
  0 & 0 & 0
 \end{pmatrix}
 \begin{pmatrix}
  x^2\\
  x\\
  1
 \end{pmatrix},
\end{align}
\begin{align}
 \begin{pmatrix}
  H_1(x) \\
  H_2(x) \\
  H_3(x) \\
 \end{pmatrix}
 = &
 \begin{pmatrix}
      0 & 0 & a_0^2(a_0^2-a_1^2)\\
      0 & (a_0^2-a_1^2)(2a_0^2-a_1^2) & -a_0(2a_0^2c_1 - a_1^2c_1 + a_0a_1c_0)\\
  a_0^2(a_0^2-a_1^2) & -a_0(2a_0^2c_1 - a_1^2c_1 + a_0a_1c_0) & a_0c_1(a_0c_1+a_1c_0)\\
 \end{pmatrix}
 \begin{pmatrix}
  x^2\\
  x\\
  1
 \end{pmatrix}\nonumber\\
 & \times
 \begin{pmatrix}
  a_0a_1 &  a_0c_0 & 0 \\
  a_0c_0 & -c_0c_1 & 0 \\
  0 & 0 & 0
 \end{pmatrix}
 \begin{pmatrix}
  x^2\\
  x\\
  1
 \end{pmatrix},
\end{align}

\subsection{Y8}
Introducing a dependent variable transformation $x_n = iy_n,\, a_1' =
ia_1$ to Eq.~\eqref{Y8}
\begin{equation*}
 x_{n+2}x_{n-1} = \frac{a_0+a_1x_n}{-a_1x_n+a_0x_nx_{n+1}},\tag{Y8}
\end{equation*}
we obtain
\begin{equation*}
 y_{n+2}y_{n-1} = \frac{a_0+a_1'y_n}{a_1'y_n+a_0y_ny_{n+1}}.
\end{equation*}
This is equivalent to Eq.~\eqref{Y7}.

\subsection{Y9}
Introducing a variable transformation $x_n=\dfrac{\widetilde
f_{n+1}}{\widetilde f_n}$ to Eq.~\eqref{Y9}
\begin{equation} x_{n+2}x_{n-1} =
\frac{x_n+x_nx_{n+1}}{1+x_n}, \tag{Y9}
\end{equation}
we obtain
\begin{equation}
\frac{\widetilde f_{n+3}\widetilde f_n}{\widetilde f_{n+2}+\widetilde
 f_{n+1}} = \frac{\widetilde f_{n+2}\widetilde f_{n-1}}{\widetilde f_{n+1}+\widetilde f_n}.
\end{equation}
From this equation, we obtain
\begin{equation}
\widetilde f_{n+2}\widetilde f_{n-1} = \alpha(\widetilde f_{n+1}+\widetilde f_n) 
\end{equation}
By scaling as $\widetilde f_n = \alpha f_n$, we obtain
\begin{equation}
f_{n+2}f_{n-1} = f_{n+1}+f_n
\end{equation}
This is Eq.~\eqref{mY6} in the case of $a=0$.

\subsection{Y2}
Eq.~\eqref{Y2} is written as
\begin{equation}  \label{mY2}
  x_{n+2}x_{n-1}=\frac{a+ax_n+ax_{n+1}+x_nx_{n+1}}{a+x_n+x_{n+1}+x_nx_{n+1}} \tag{Y2}
\end{equation}
where $a = a_0/a_3$. Eq.~\eqref{mY2} is generated by a pair of
second-order equation
\begin{equation}\label{Y2second}
 \begin{cases}
  g_{n+1}=-g_n\left(1+\dfrac{(a+g_{n-1})(b_0(a-g_n^2)-ac)}{(a-1)^2(a-g_n)(g_n+g_{n-1})+b_0g_n(a+ag_n+ag_{n-1}+g_ng_{n-1})-acg_n}\right),\\
  \vbox to 8mm{}
  h_{n+1}=-h_n\left(1+\dfrac{(a+h_{n-1})(b_1(a-h_n^2)-ac)}{(a-1)^2(a-h_n)(h_n+h_{n-1})+b_1g_n(a+ah_n+ah_{n-1}+h_nh_{n-1})-ach_n}\right),
 \end{cases}
\end{equation}
where $g_n = x_{2n},\ h_n = x_{2n+1}$ and
\begin{align}
  c &= \frac{1}{x_0x_1x_2}(1 + x_0)(1 + x_1)(1 + x_2)(a(1 + x_0 + x_1 + x_2) + x_0x_1 + x_1x_2 + x_2x_0 + x_0x_1x_2), \\
  b_0 &= \frac{1}{x_0x_1x_2}\{(a-1)x_0x_2(1+x_1)^2 \nonumber\\
      &\qquad + (1+x_1 + x_0x_1 + x_1x_2)(a(1 + x_0 + x_1 + x_2) + x_0x_1 + x_1x_2 + x_2x_0 + x_0x_1x_2),\\
  b_1 &= \frac{1}{x_1x_2x_3}\{(a-1)x_1x_3(1+x_2)^2 \nonumber\\
      &\qquad + (1+x_2 + x_1x_2 + x_2x_3)(a(1 + x_1 + x_2 + x_3) + x_1x_2 + x_2x_3 + x_3x_1 + x_1x_2x_3).
\end{align}

Eq.~\eqref{Y2second} does not belong to the QRT system as it stands,
though there is still a possibility that it could be reduced to the
system through a change of variable.

\subsection{Y3}
About Eq.~\eqref{Y3}, we can only show numerical results. For various
parameter $a_0,\ a_3 $ and initial values $x_0\sim x_2$, it is generated
by a pair of second-order equations on even steps ($g_n$) and odd
steps ($h_n$). In any case, both equations follows
\begin{equation}\label{Y3second}
 g_{n+1} = \frac{\displaystyle\sum_{0\le i,j \le 3}a_{ij}g_{n-1}^ig_n^j}{\displaystyle\sum_{0\le i,j \le 3}b_{ij}g_{n-1}^ig_n^j}, \quad  h_{n+1} = \frac{\displaystyle\sum_{0\le i,j \le 3}c_{ij}h_{n-1}^ih_n^j}{\displaystyle\sum_{0\le i,j \le 3}d_{ij}h_{n-1}^ih_n^j}, 
\end{equation}
where $a_{ij}\sim d_{ij}$ are constant obtained from parameters and
initial values numerically . This fact strongly suggests that
Eq.~\eqref{Y3} is also derived from a pair of second-order equations.

\section{ultradiscretization}
In this section, we consider an ultradiscrete version of the third-order
integrable equations\cite{TTMS,MSTTT}. Since ultradiscretization
requires a positivity of parameters and dependent variables of the
equations, (Y1),(Y2),(Y4),(Y6),(Y7) and (Y9) are ultradiscretizable. We
show the procedure of ultradiscretization works well by Eq.~\eqref{mY6}
as an example.

\begin{figure}[htbp]
 \includegraphics{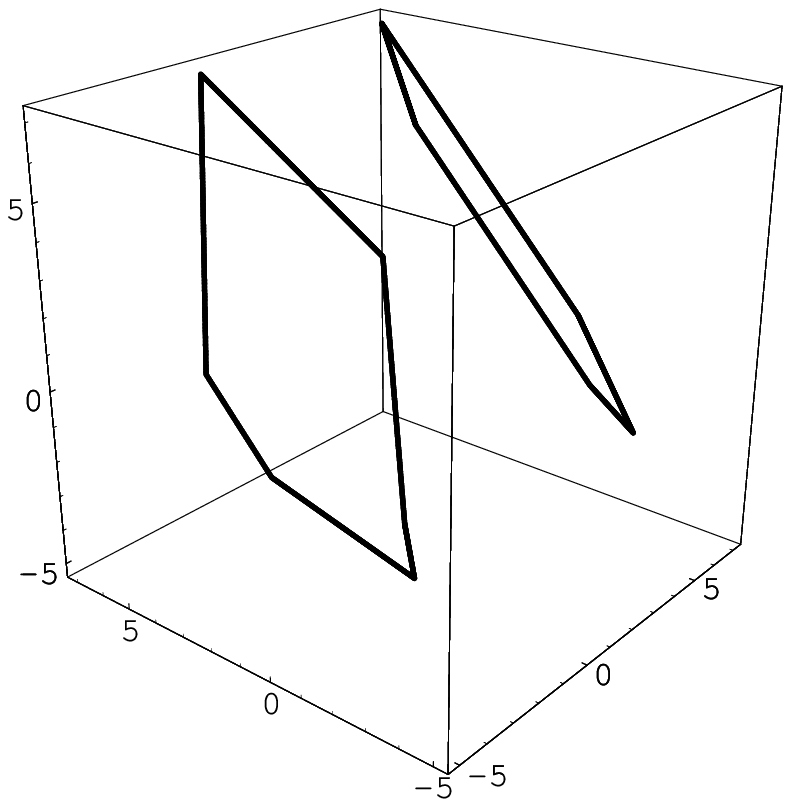} \caption{ \label{fig:fig2}A trajectory of a
 solution to Eq.~\eqref{umY6} for $A=2.0,\, Y_0=2.0,\, Y_1=2.01,\,
 Y_2=-5.1$}
\end{figure}

If we use transformations $y_n =\exp\left(\dfrac{Y_n}{\epsilon}\right),\
a = \exp\left(\dfrac{A}{\epsilon}\right)$ for Eq.~\eqref{mY6}
\begin{equation*}
 y_{n+2}y_{n-1} = a + y_n + y_{n+1},
\end{equation*}
and take a limit $\epsilon\rightarrow +0$, then we have
\begin{equation}\label{umY6}
 Y_{n+2} = \max (A, Y_n, Y_{n+1}) - Y_{n-1}.
\end{equation}
This is an ultradiscrete version of Eq.~\eqref{mY6}. Fig.\ref{fig:fig2}
shows a trajectory of a solution to Eq.~\eqref{umY6} in 3D phase space.

\begin{figure}[htbp]
 \includegraphics{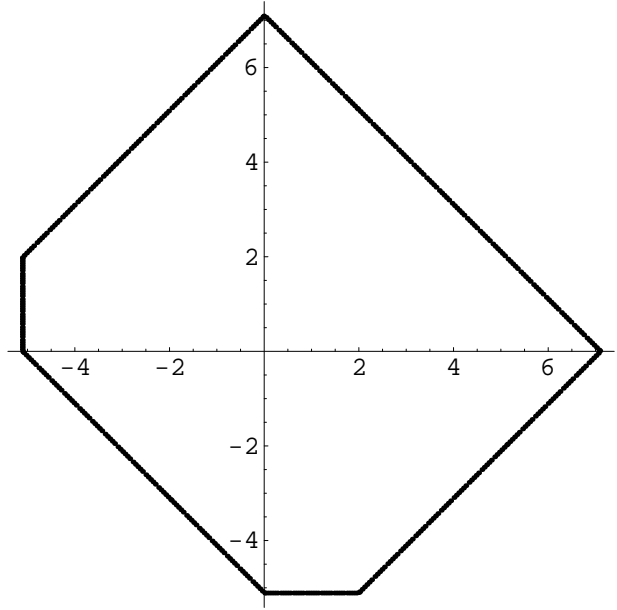} 
 \caption{ \label{fig:fig3}An invariant curve
 of Eq.~\eqref{umY6U} for $A=2.0,\, Y_0=2.0,\, Y_1=2.01,\, Y_2=-5.1$}
\end{figure}

It follows from the result for Eq.~\eqref{mY6} in the previous section
that Eq.~\eqref{umY6} is generated by a pair of ultradiscrete QRT
system
\begin{align}
 U_{n+1} &= \max (0, A+C_0, U_n + \max(0, C_0)) - U_{n-1} - \max(0,
 U_n), \label{umY6U}\\
 V_{n+1} &= \max (0, A+C_1, V_n + \max(0, C_1)) - V_{n-1} - \max(0, V_n),\\
 C_0 &= \max(0, Y_0) + \max(0, Y_2) - Y_1,\\
 C_1 &= \max(0, Y_1) + \max(0, Y_3) - Y_2,
\end{align}
and invariant curves become
\begin{align}\label{umY6i}
\max\Bigl(&U_n+U_{n+1},\ \max(U_n, U_{n+1}) + \max(0, C_0),\ \max(U_n
 -U_{n+1}, U_{n+1} - U_n) + \max(0, C_0),\ \nonumber\\ & \max(-U_n,
 -U_{n+1}) + \max(0, C_0, A+C_0, 2C_0),\nonumber\\
& - U_n - U_{n+1} + \max(0, C_0) + \max(0,
 A+C_0) \Bigr) = K_0,
\end{align}
and
\begin{align}
\max\Bigl(&V_n+V_{n+1},\ \max(V_n, V_{n+1}) + \max(0, C_1),\ \max(V_n
 -V_{n+1}, V_{n+1} - V_n) + \max(0, C_1),\ \nonumber\\ & \max(-V_n,
 -V_{n+1}) + \max(0, C_1, A+C_1, 2C_1),\nonumber\\
& - V_n - V_{n+1} + \max(0, C_1) + \max(0, A+C_1) \Bigr) = K_1
\end{align}
Fig.\ref{fig:fig3} shows invariant curves for Eq.~\eqref{umY6U}
determined by Eq.~\eqref{umY6i}.

\section{conclusion}
In this paper, we have shown the third-order integrable difference
equations proposed by Hirota, Kimura and Yahagi are generated by a pair
of second-order integrable difference equations. In the case of
Eqs.~\eqref{Y1} and \eqref{Y4}$\sim$\eqref{Y9}, second-order difference
equations are a special case of the QRT system. In the case of
Eqs.~\eqref{Y2} and \eqref{Y3}, second-order equations may not be the
QRT system. Furthermore we have shown the procedure of
ultra-discretization works well for third-order equation, and derived
second-order equations and invariants curves are also ultradiscretizable
.

Although the whole integrability structure of the general third-order
equations is still unknown, our work would be one of keys to understand
the structure. Generating our results, that is, investigating a
connection between the general QRT system and the third-order equations
is an important future problem.

\begin{acknowledgments}
 The authors express our sincere thanks to Professor Ryogo Hirota for
 fruitful discussions and encouragement.
\end{acknowledgments}
\bibliography{third}

\begin{thebibliography}{14}
\expandafter\ifx\csname natexlab\endcsname\relax\def\natexlab#1{#1}\fi
\expandafter\ifx\csname bibnamefont\endcsname\relax
  \def\bibnamefont#1{#1}\fi
\expandafter\ifx\csname bibfnamefont\endcsname\relax
  \def\bibfnamefont#1{#1}\fi
\expandafter\ifx\csname citenamefont\endcsname\relax
  \def\citenamefont#1{#1}\fi
\expandafter\ifx\csname url\endcsname\relax
  \def\url#1{\texttt{#1}}\fi
\expandafter\ifx\csname urlprefix\endcsname\relax\def\urlprefix{URL }\fi
\providecommand{\bibinfo}[2]{#2}
\providecommand{\eprint}[2][]{\url{#2}}

\bibitem[{\citenamefont{Grammaticos et~al.}(1991)\citenamefont{Grammaticos,
  Ramani, and Papageorgiou}}]{GRP}
\bibinfo{author}{\bibfnamefont{B.}~\bibnamefont{Grammaticos}},
  \bibinfo{author}{\bibfnamefont{A.}~\bibnamefont{Ramani}}, \bibnamefont{and}
  \bibinfo{author}{\bibfnamefont{V.}~\bibnamefont{Papageorgiou}},
  \bibinfo{journal}{Phys.\ Rev.\ Lett.} \textbf{\bibinfo{volume}{67}},
  \bibinfo{pages}{1825} (\bibinfo{year}{1991}).

\bibitem[{\citenamefont{Hietarinta and Viallet}(1998)}]{HV}
\bibinfo{author}{\bibfnamefont{J.}~\bibnamefont{Hietarinta}} \bibnamefont{and}
  \bibinfo{author}{\bibfnamefont{C.}~\bibnamefont{Viallet}},
  \bibinfo{journal}{Phys.\ Rev.\ Lett.} \textbf{\bibinfo{volume}{81}},
  \bibinfo{pages}{325} (\bibinfo{year}{1998}).

\bibitem[{\citenamefont{Sakai}(2001)}]{Sakai}
\bibinfo{author}{\bibfnamefont{H.}~\bibnamefont{Sakai}},
  \bibinfo{journal}{Commun.\ Math.\ Phys.} \textbf{\bibinfo{volume}{220}},
  \bibinfo{pages}{165} (\bibinfo{year}{2001}).

\bibitem[{\citenamefont{Takenawa}(2001)}]{Take}
\bibinfo{author}{\bibfnamefont{T.}~\bibnamefont{Takenawa}},
  \bibinfo{journal}{J.\ Phys.\ A.} \textbf{\bibinfo{volume}{34}},
  \bibinfo{pages}{L95} (\bibinfo{year}{2001}).

\bibitem[{\citenamefont{Ramani et~al.}(2001{\natexlab{a}})\citenamefont{Ramani,
  Grammaticos, and Ohta}}]{RGO1}
\bibinfo{author}{\bibfnamefont{A.}~\bibnamefont{Ramani}},
  \bibinfo{author}{\bibfnamefont{B.}~\bibnamefont{Grammaticos}},
  \bibnamefont{and} \bibinfo{author}{\bibfnamefont{Y.}~\bibnamefont{Ohta}},
  \bibinfo{journal}{Commun.\ Math.\ Phys.} \textbf{\bibinfo{volume}{217}},
  \bibinfo{pages}{315} (\bibinfo{year}{2001}{\natexlab{a}}).

\bibitem[{\citenamefont{Ramani et~al.}(2001{\natexlab{b}})\citenamefont{Ramani,
  Grammaticos, and Ohta}}]{RGO2}
\bibinfo{author}{\bibfnamefont{A.}~\bibnamefont{Ramani}},
  \bibinfo{author}{\bibfnamefont{B.}~\bibnamefont{Grammaticos}},
  \bibnamefont{and} \bibinfo{author}{\bibfnamefont{Y.}~\bibnamefont{Ohta}},
  \bibinfo{journal}{J.\ Phys.\ A.} \textbf{\bibinfo{volume}{34}},
  \bibinfo{pages}{2505} (\bibinfo{year}{2001}{\natexlab{b}}).

\bibitem[{\citenamefont{Tsuda}(2004)}]{Tsuda}
\bibinfo{author}{\bibfnamefont{T.}~\bibnamefont{Tsuda}}, \bibinfo{journal}{J.\
  Phys.\ A.} \textbf{\bibinfo{volume}{37}}, \bibinfo{pages}{2721}
  (\bibinfo{year}{2004}).

\bibitem[{\citenamefont{Kajiwara et~al.}(2003)\citenamefont{Kajiwara, Masuda,
  Noumi, Ohta, and Yamada}}]{KMNOY}
\bibinfo{author}{\bibfnamefont{K.}~\bibnamefont{Kajiwara}},
  \bibinfo{author}{\bibfnamefont{T.}~\bibnamefont{Masuda}},
  \bibinfo{author}{\bibfnamefont{M.}~\bibnamefont{Noumi}},
  \bibinfo{author}{\bibfnamefont{Y.}~\bibnamefont{Ohta}}, \bibnamefont{and}
  \bibinfo{author}{\bibfnamefont{Y.}~\bibnamefont{Yamada}},
  \bibinfo{journal}{J.\ Phys.\ A.} \textbf{\bibinfo{volume}{36}},
  \bibinfo{pages}{L263} (\bibinfo{year}{2003}).

\bibitem[{\citenamefont{Quispel et~al.}(1988)\citenamefont{Quispel, Robert, and
  Thompson}}]{QRT1}
\bibinfo{author}{\bibfnamefont{G.}~\bibnamefont{Quispel}},
  \bibinfo{author}{\bibfnamefont{J.}~\bibnamefont{Robert}}, \bibnamefont{and}
  \bibinfo{author}{\bibfnamefont{C.}~\bibnamefont{Thompson}},
  \bibinfo{journal}{Phys.\ Lett.\ A.} \textbf{\bibinfo{volume}{126}},
  \bibinfo{pages}{419} (\bibinfo{year}{1988}).

\bibitem[{\citenamefont{Quispel et~al.}(1989)\citenamefont{Quispel, Robert, and
  Thompson}}]{QRT2}
\bibinfo{author}{\bibfnamefont{G.}~\bibnamefont{Quispel}},
  \bibinfo{author}{\bibfnamefont{J.}~\bibnamefont{Robert}}, \bibnamefont{and}
  \bibinfo{author}{\bibfnamefont{C.}~\bibnamefont{Thompson}},
  \bibinfo{journal}{Physica.\ D} \textbf{\bibinfo{volume}{34}},
  \bibinfo{pages}{183} (\bibinfo{year}{1989}).

\bibitem[{\citenamefont{Ramani et~al.}(1991)\citenamefont{Ramani, Grammaticos,
  and Hietarinta}}]{RGH}
\bibinfo{author}{\bibfnamefont{A.}~\bibnamefont{Ramani}},
  \bibinfo{author}{\bibfnamefont{B.}~\bibnamefont{Grammaticos}},
  \bibnamefont{and}
  \bibinfo{author}{\bibfnamefont{J.}~\bibnamefont{Hietarinta}},
  \bibinfo{journal}{Phys.\ Rev.\ Lett.} \textbf{\bibinfo{volume}{67}},
  \bibinfo{pages}{1829} (\bibinfo{year}{1991}).

\bibitem[{\citenamefont{Hirota et~al.}(2001)\citenamefont{Hirota, Kimura, and
  Yahagi}}]{HKY}
\bibinfo{author}{\bibfnamefont{R.}~\bibnamefont{Hirota}},
  \bibinfo{author}{\bibfnamefont{K.}~\bibnamefont{Kimura}}, \bibnamefont{and}
  \bibinfo{author}{\bibfnamefont{H.}~\bibnamefont{Yahagi}},
  \bibinfo{journal}{J.\ Phys.\ A.} \textbf{\bibinfo{volume}{34}},
  \bibinfo{pages}{10377} (\bibinfo{year}{2001}).

\bibitem[{\citenamefont{Tokihiro et~al.}(1996)\citenamefont{Tokihiro,
  Takahashi, Matsukidaira, and Satsuma}}]{TTMS}
\bibinfo{author}{\bibfnamefont{T.}~\bibnamefont{Tokihiro}},
  \bibinfo{author}{\bibfnamefont{D.}~\bibnamefont{Takahashi}},
  \bibinfo{author}{\bibfnamefont{J.}~\bibnamefont{Matsukidaira}},
  \bibnamefont{and} \bibinfo{author}{\bibfnamefont{J.}~\bibnamefont{Satsuma}},
  \bibinfo{journal}{Phys.\ Rev.\ Lett.} \textbf{\bibinfo{volume}{76}},
  \bibinfo{pages}{3247} (\bibinfo{year}{1996}).

\bibitem[{\citenamefont{Matsukidaira et~al.}(1997)\citenamefont{Matsukidaira,
  Satsuma, Takahashi, Tokihiro, and Torii}}]{MSTTT}
\bibinfo{author}{\bibfnamefont{J.}~\bibnamefont{Matsukidaira}},
  \bibinfo{author}{\bibfnamefont{J.}~\bibnamefont{Satsuma}},
  \bibinfo{author}{\bibfnamefont{D.}~\bibnamefont{Takahashi}},
  \bibinfo{author}{\bibfnamefont{T.}~\bibnamefont{Tokihiro}}, \bibnamefont{and}
  \bibinfo{author}{\bibfnamefont{M.}~\bibnamefont{Torii}},
  \bibinfo{journal}{Phys.\ Lett.\ A.} \textbf{\bibinfo{volume}{255}},
  \bibinfo{pages}{287} (\bibinfo{year}{1997}).

\end{thebibliography}

\end{document}